\documentclass{aaarxiv}
\usepackage[varg]{txfonts}
\usepackage{natbib}
\usepackage[dvipsnames]{xcolor}
\usepackage{soul}
\usepackage{algorithm2e}
\usepackage{hyperref}

\newcommand{\mhz}{\mathrm{\mu Hz}}
\newcommand{\fap}{$w$,$A$,$\delta$}

\begin{document}
\title{LASR-Guided Stellar Photometric Variability Subtraction: The Linear Algorithm For Significance Reduction}
\titlerunning{The Linear Algorithm for Significance Reduction}

\author{John P. Ahlers\inst{1}
	\and Jason W. Barnes\inst{1} 
	\and Sarah A. Horvath\inst{1}
	\and Samuel A. Myers\inst{1}
	\and Matthew M. Hedman\inst{1}} 

\institute{Department of Physics, University of Idaho, 875 Perimeter Drive MS 0903, Moscow, ID 83844, USA} 

\abstract{We develop a technique for removing stellar variability in the light curves of $\delta$-Scuti and similar stars. Our technique, which we name the Linear Algorithm for Significance Reduction (LASR), subtracts oscillations from a time series by minimizing their statistical significance in frequency space. We demonstrate that LASR can subtract variable signals of near-arbitrary complexity and can robustly handle close frequency pairs and overtone frequencies. We demonstrate that our algorithm performs an equivalent fit as prewhitening to the straightforward variable signal of KIC 9700322. We also show that LASR provides a better fit to seismic activity than prewhitening in the case of the complex $\delta$-Scuti KOI-976.}

\maketitle 

\section{Introduction}
Recent observing programs such as the \emph{Kepler} mission have demonstrated that stellar variability in high mass stars often renders transit light curves unusable. \citep{basri2010photometric}. Additionally, high-mass stars often rotate rapidly, inducing an oblate shape and a pole-to-equator luminosity gradient across the stellar surface \citep{barnes2009transit,barnes2011measurement,ahlers2015spin,ahlers2016gravity}. These two effects add challenges to traditional light curve analysis, radial velocity measurements, Doppler tomography, and Rossiter McLaughlin measurements \citep{udry2007decade,gimenez2006equations}.

Approximately 60\% of stars in the \emph{Kepler} field of view display more stellar variability than the Sun \citep{mcquillan2013stellar}. Such variability produces effects in transit light curves that make traditional fitting challenging or impossible. Many of these targets are low-mass stars with variability caused by non-sinusoidal effects such as sunspots in their convective exteriors. Techniques for analyzing non-sinusoidal or non-periodic signals in the light curves such as the autocorrelation function \citep{mcquillan2014rotation} and Gaussian processes \citep{aigrain2016k2sc} produce strong results when applied to such stars. However, high-mass stars behave quite differently. At $\sim6250$K and hotter, stars invert to become radiative rather than convective at their surface \citep{winn2010hot}. These stars have weak or nonexistent sunspots, and commonly rotate rapidly as a mostly-rigid body throughout their lifetimes. High-mass stars in the classical instability strip pulsate with radial and nonradial modes at high amplitudes. Therefore, analysis of stellar variability in the light curves of high-mass stars comes with a unique set of challenges and must be handled differently than variability in low-mass stars.    

For classical pulsators such as Delta Scuti ($\delta$-Scuti) and Gamma Doradus stars, the technique of ``prewhitening" serves as the traditional method of asteroseismic analysis of transit light curves \citep[e.g.,][]{hernandez2009asteroseismic,poretti2009hd}. Prewhitening fits sinusoids to photometry in an iterative process. This algorithm performs least-squares fits of several of the highest-amplitude oscillations in the time domain, determining approximate frequency, amplitude, and phase values for those oscillations. Prewhitening then fits several next-highest amplitude oscillations as a running total with the original fit, repeating this process until stellar variability has been resolved.   

Prewhitening often serves as an adequate method for removing stellar variability and determining the frequencies of oscillation in classical pulsators. However, we explore an alternate route out of necessity: we found that prewhitening provided an inadequate fit of the complex signal of Kepler Object of Interest (KOI) 976. KOI-976 is a high-amplitude, rapidly rotating $\delta$-Scuti star with a complex variable signal. We tried to remove stellar variability from the transit light curve of KOI-976 by applying prewhitening, but we found that the complex signal contained too many oscillations for an accurate least-squares fit in the time domain. This roadblock led us to explore removing stellar variability in frequency-space instead through a new process that we call the Linear Algorithm for Significance Reduction (LASR). 

In this paper we develop a frequency-domain method for removing the asteroseismic signal of high-mass pulsators. In $\S$\ref{sec:methods} we detail the LASR technique. In $\S$\ref{sec:results} we apply LASR to a synthetic dataset and to $\delta$-Scutis KIC 9700322 and KOI-976 to compare our technique against prewhitening. In $\S$4 we discuss how LASR compares with existing techniques for analyzing variability in photometric data.

\section{Methods}\label{sec:methods}
The Linear Algorithm for Significance Reduction (LASR) serves as an alternate method to prewhitening for signal reduction. It resolves a linear combination of oscillations in a time series by minimizing each oscillation's significance in the frequency domain. The algorithm operates linearly: it reduces the highest-amplitude frequency, subtracts it from the time series, and then reduces the new highest-amplitude frequency. 

The LASR technique has two primary advantages over traditional prewhitening. First, because it operates in frequency space, LASR's fitting process treats every oscillation as independent and avoids the degeneracies and complex parameter space that prewhitening encounters for datasets containing many oscillation modes. Second, the computer code behind LASR is extremely simple to run and requires very little knowledge of signal processing, making it an accessible technique for inexperienced researchers. In $\S$\ref{sec:algorithm} we detail the algorithm for significance reduction. In $\S$\ref{sec:interdependent} we discuss handling interdependent frequencies including close frequency pairs and integer-multiple frequencies. In $\S$\ref{sec:error} and $\S$\ref{sec:error}, we detail our best-fit error analysis. In Appendix \ref{app:deriv} we mathematically derive that LASR's straight-forward approach to subtracting stellar oscillations is robust for sinusoidal variability.

\subsection{LASR Algorithm} \label{sec:algorithm}
\begin{figure}[th]
\includegraphics{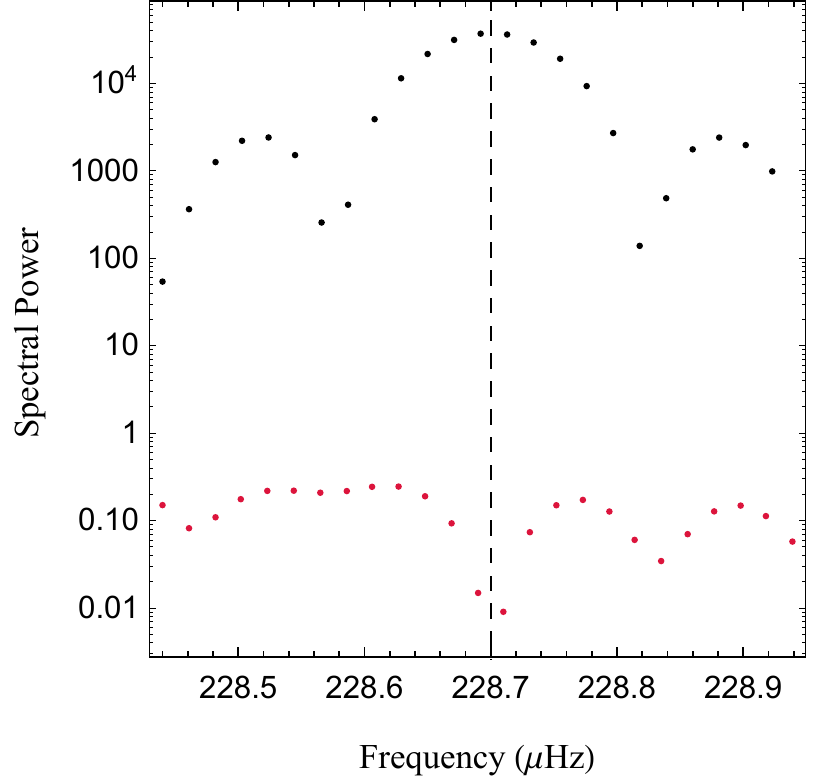}
\caption{\footnotesize A periodogram window for the highest-amplitude peak in the synthetic time series discussed in $\S$\ref{sec:synthetic}. We set LASR to sample the spectral power of this peak 25 times. The black and red points show the $P_i(\omega)$ before and after LASR minimized the peak's significance. The unreduced oscillation shows a clear peak with aliasing on either side. The dashed line marks the true frequency of this oscillation. LASR successfully reduces this oscillation and yields correct measurements of its (\fap) values (see Table \ref{table:synthresults}.)} 
\label{fig:synthpeak}
\end{figure}

LASR combines two well-known tools for signal processing: the Lomb-Scargle normalized periodogram, and the downhill simplex routine. To remove a single oscillation from a time series, LASR creates a window of the power spectrum of the time series around the peak of that oscillation (Figure \ref{fig:synthpeak}). It then applies the downhill simplex routine to minimize that peak's significance. 

We use the traditional Lomb-Scargle normalized periodogram \citep{press2007numerical} for spectral analysis. The variations in brightness in high-amplitude $\delta$-Scuti stars correspond to $\sim$5\% variations in uncertainty. We test whether these variations in uncertainty affect LASR by comparing the traditional Lomb-Scargle normalized periodogram with the generalized periodogram \citep[e.g.,][]{zechmeister2009generalised,vio2010unevenly}, which applies weights to each time bin according to its photometric uncertainty. We find no noticeable differences between the two methods for $\delta$-Scuti stars KOI-976 and KIC 9700322. In the case of different noise properties and instrument systematics, more computationally expensive techniques that better handle time bin uncertainty may provide better results in the significance reduction process. Such systematics do not appear in this analysis; we therefore favor the traditional Lomb-Scargle periodogram,
\begin{equation}
P(\omega) =\frac{1}{2\sigma^2}\sum_{\phi=0,\frac{\pi}{2}} \frac{|\sum_j(h_j-\bar{h})\sin(\omega(t_j-\tau)+\phi)|^2}{\sum_j\sin^2(\omega(t_j-\tau)+\phi)}
\end{equation}

where $h_j$ is the photometric flux value at time $t_j$, $\sigma$ is the standard deviation in the dataset, $\phi$ is a phase offset that includes the values $0$ and $\pi/2$, and $\omega=2\pi f$ is the angular frequency being sampled. The offset $\tau$ is defined as,
\begin{equation}
\tan(2\omega\tau) = \frac{\sum_j\sin(2\omega t_j)}{\sum_j\cos(2\omega t_j)} 
\end{equation} 

Several works exist to explain the statistical significance of frequency peaks in a periodogram \citep[e.g.,][]{press2007numerical,baluev2008assessing}. We represent the statistical significance of an oscillation in our dataset by sampling a window of frequencies around the peak. LASR evenly samples frequencies in a window approximately three times the full width of the peak.

LASR's adjustable window width depends on the width of frequency peaks; in general, a longer time series means narrower peaks in a periodogram. For KOI-976's short cadence photometry discussed in $\S$3, we sample $P(\omega)$ 40 times in a window width of $0.7\mhz$. We sum $P(\omega)$ values and use the resulting value to represent the significance $S_i(\omega,A,\delta)$ of the $i^{th}$ oscillation in our dataset as a function of frequency $\omega$, amplitude $A$, and phase $\delta$. 

The goal of LASR is to find the (\fap) values that minimize $S(\omega,A,\delta)$. To do this, LASR uses a downhill simplex routine \citep{nelder1965simplex,press2007numerical} to find the minimum of $S(\omega,A,\delta)$. We choose this minimization routine over more robust routines such as Powell's Method \citep{brent2013algorithms} because of the well-behaved parameter space that results from minimizing oscillations in the frequency domain (see Appendix \ref{app:deriv}). In Appendix \ref{app:inputs} we list the necessary inputs to operate LASR and discuss computation times for running the algorithm and in \ref{app:code} we provide pseudocode for writing this routine in any computer programming language. 

\subsection{LASR and Interdependent Frequencies} \label{sec:interdependent}
In general, LASR removes oscillations from a dataset one at a time because they are linearly independent from one another. However, two scenarios arise where this assumption fails: close frequency pairs and overtone frequencies. 

\begin{figure}[th]
\includegraphics{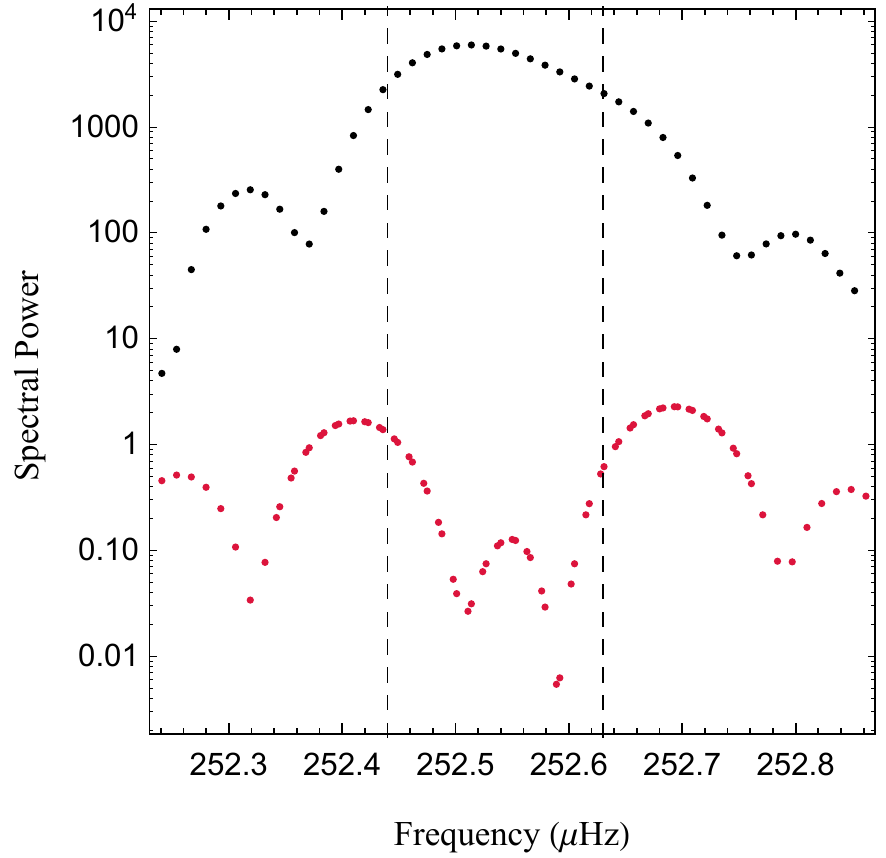}
\caption{\footnotesize Log-scale plot of the power spectrum of two close frequencies ($f_2$ and $f_3$ in Table \ref{table:synthresults}). The two oscillations appear in frequency space as a single asymmetric peak (black). LASR minimizes the significance of both peaks simultaneously (red) and accurately determines the (\fap) of both oscillations. LASR samples frequencies of a periodogram window for each frequency; for close frequency pairs these two windows overlap.}
\label{fig:synthclosefreqs}
\end{figure}

We define close frequency pairs as oscillations close enough together in frequency space that spectral analysis cannot resolve them individually (Figure \ref{fig:synthclosefreqs}), which can cause problems for prewhitening. If two frequencies exist close enough together that their periodogram windows (described in $\S$\ref{sec:algorithm}) overlap, then $S_i(\omega,A,\delta)$ cannot be properly minimized. Close frequency pairs are common both for complex datasets with many independent oscillations and for short time series that yield wide frequency peaks in their periodograms. LASR's solution, however, is simple: just remove both peaks at once. LASR can remove any number of peaks simultaneously by minimizing a combined $S_i(\omega,A,\delta)$ function that samples $P(\omega)$ values for all relevant peaks, and then minimizing that function within a single large simplex.

Overtone frequencies can be more challenging because they are not mathematically independent. The power spectrum of an oscillation can be changed dramatically by integer multiple frequencies. Therefore, minimizing $S_i(\omega,A,\delta)$ for a single frequency in this scenario will result in a poor determination of its (\fap) values.

\renewcommand{\arraystretch}{1.2}
\
\begin{table}[h]
\centering 
\begin{tabular}{p{0.28\textwidth}p{0.1\textwidth}}
\hline
\hline
{\bf Synthetic Time Series Quantity} & {\bf Value}  \\ \hline
Length of synthetic dataset & $90~\mathrm{days}$ \\ 
Photometric cadence  & $1~\mathrm{min}$ \\
Flux normalization constant & $1.0$ \\
Transit gap start times & $\sum_{n=1}^4 20n~\mathrm{days}$ \\
Transit gap lengths & $10~\mathrm{hr}$ \\
Large gap start time & $4.5\times10^6~\mathrm{s}$ \\
Large gap length & $3.0\times10^5~\mathrm{s}$ \\
Gaussian uncertainty & $5.0\times10^{-4}$ \\
Lag correlation coefficient & $0.5$ \\
Injected transit depth & $1.2\times10^{-4}$ \\
Injected transit period & $15~\mathrm{days}$ 
\end{tabular}
\caption{\footnotesize Global parameters of the synthetic data we generate to test LASR. We choose these parameters based on typical quantities of \emph{Kepler} short-cadence photometry. The lag correlation coefficient sets the lag-1 autocorrelation between successive time samples, transforming Gaussian noise to correlated noise \citep{haykin2006nonlinear}.}
\label{table:synthparams}
\end{table}

\begin{figure*}[th]
\includegraphics[width=\textwidth]{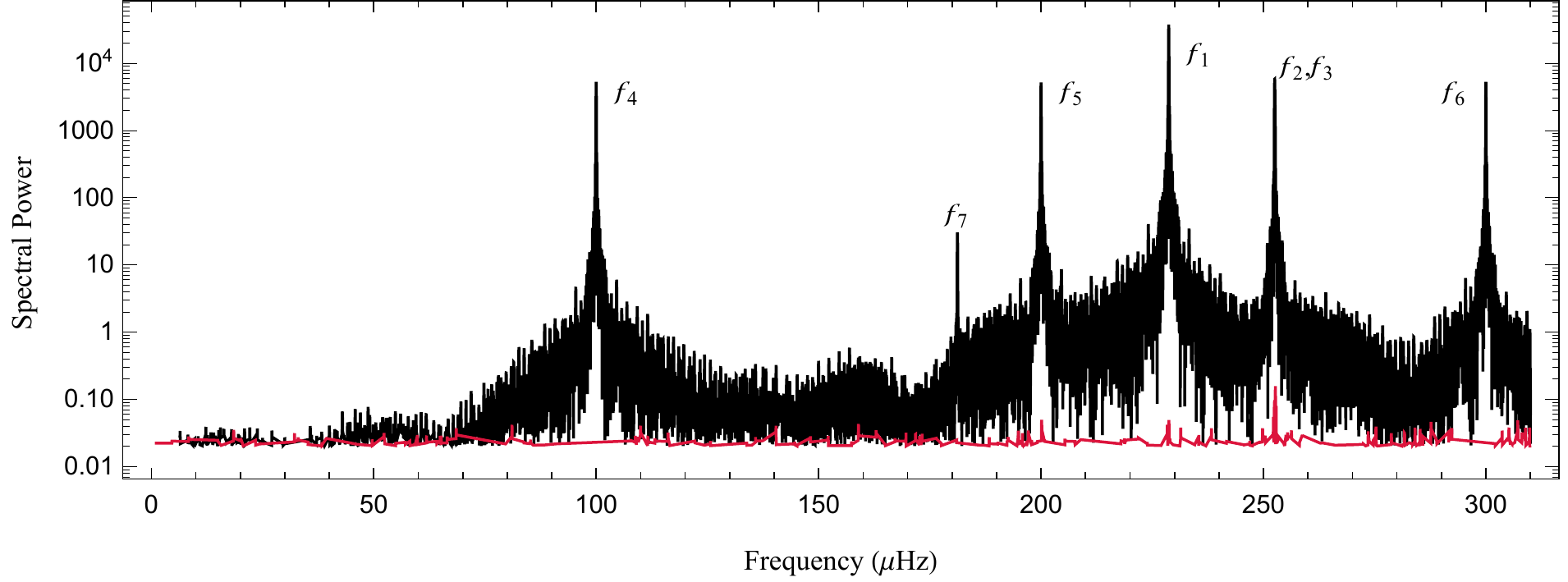}
\caption{\footnotesize Total power spectrum of the synthetic time series before (black) and after (red) LASR reduction of its seven oscillations. As we show in Table \ref{table:synthresults}, LASR reduces all oscillations by over 99.9\% of their original significance and yields accurate (\fap) values. The injected transit causes no noticeable effect on the power spectrum.}
\label{fig:synthperiodo}
\end{figure*}

\renewcommand{\arraystretch}{1.5}
\begin{table*}[htp] 
\centering
\noindent\begin{tabular}{l|lllllllll}
\hline
\hline

$f_\mathrm{\#}$ &
\begin{tabular}{l}\hspace{-0.75em}$f_\mathrm{start}$ \\\hspace{-0.75em}($\mhz$) \end{tabular} & 
\begin{tabular}{l}\hspace{-0.75em}$f_\mathrm{result}$ \\\hspace{-0.75em}($\mhz$) \end{tabular} &
\begin{tabular}{l}\hspace{-0.75em}$f_\mathrm{actual}$ \\\hspace{-0.75em}($\mhz$) \end{tabular} &
\begin{tabular}{l}\hspace{-0.75em}$a_\mathrm{start}$ \\\hspace{-0.75em}($\mathrm{10^{-3}}$) \end{tabular} & 
\begin{tabular}{l}\hspace{-0.75em}$a_\mathrm{result}$ \\\hspace{-0.75em}($\mathrm{10^{-3}}$) \end{tabular} &
\begin{tabular}{l}\hspace{-0.75em}$a_\mathrm{actual}$ \\\hspace{-0.75em}($\mathrm{10^{-3}}$) \end{tabular} &
$\delta_\mathrm{start}$ & $\delta_\mathrm{result}$ & $\delta_\mathrm{actual}$ \\ \hline
1 & 228.68 & 228.7$\pm$5e-6 & 228.7 & 10.0 &  20.0013$\pm$0.0014 & 20.0 & 3.000 & 3.99998$\pm$0.00014 & 4.000 \\
2 & 252.44 & 252.5$\pm$1.1e-5 & 252.5 & 1.0 & 7.5026$\pm$0.0015 & 7.5 & 3.000 & 2.9838$\pm$0.0003 & 3.00 \\
3 & 252.63 & 252.6$\pm$1.7e-5 & 252.6 & 1.0 & 4.9985$\pm$0.0016 & 5.0 & 3.000 & 2.0011$\pm$0.0005 & 2.00 \\
4 & 99.930 & 100.0$\pm$1.5e-5  & 100.0 & 1.0 & 7.4997$\pm$0.0014 & 7.5 & 3.000 & 0.0006$\pm$0.0004 & 0.000 \\
5 & 199.965 & 200.0$\pm$1.3e-5 & 200.0 & 1.0 & 7.4993$\pm$0.0014 & 7.5 & 3.000 & 0.9987$\pm$0.0004 & 1.000 \\
6 & 299.97 & 300.0$\pm$1.0e-5 & 300.0 & 1.0 & 7.4998$\pm$0.0014 & 7.5 & 3.000 & 2.0009$\pm$0.0003 & 2.000 \\
7 & 181.194 & 181.1994$\pm$0.0002 & 181.2 & 0.1 & 0.4975$\pm$0.0014 & 0.5 & 3.000 & 5.013$\pm$0.006 & 5.000 \\ 
\end{tabular}
\caption{\footnotesize Oscillations added to a synthetic dataset to test LASR's ability to subtract variability. We list the seven oscillations in the order of: a single high-amplitude frequency (1), two close frequency pairs (2,3), three overtone frequencies (4,5,6), and a low-amplitude frequency whose amplitude matches the $1\sigma$ Gaussian noise of the dataset. For every oscillation, we list initial guesses, resulting best-fit values, and actual values for frequency ($f$), amplitude ($A$), and phase ($\delta$). To simulate handling a real dataset, we set our starting frequency values to the peak values measured in frequency space. We set starting amplitudes to ``reasonable guesses'' based on the photometry, and we choose the random phase value 3.0 for all oscillations.}
\label{table:synthresults}
\end{table*}

The solution is again to simply remove both peaks simultaneously. The underlying challenge stems from recognizing that this behavior is occurring in the first place. In our analysis of KOI-976, we identify all relevant frequencies through spectral analysis before applying the LASR technique to search for such resonances. We find scant evidence of this scenario arising in KOI-976, but we test simultaneous subtraction in an idealized dataset in $\S$\ref{sec:synthetic} and show its successful determination of (\fap) for overtone frequencies (Table \ref{table:synthresults}).  

\subsection{Error Analysis}\label{sec:error}
We find the uncertainty in our best-fit parameters by calculating the covariance matrix of our dataset. Following \citet{andrae2010error}, we estimate that the likelihood function $\mathcal{L}$ of our model is nearly Gaussian at its maximum, allowing us to calculate the model's covariance matrix using the Fisher information matrix $\mathcal{I}$,
\begin{equation}
\mathcal{I}_{i,j} = \left(-\frac{\partial^2\log\mathcal{L}}{\partial\theta_i\partial\theta_j} \right)
\end{equation}

where $\theta_i$ is the $i^{\mathrm{th}}$ model parameter and $\log\mathcal{L}\propto\chi^2$. We numerically approximate these second derivatives in each element of the Fisher matrix as,

\begin{equation}
\mathcal{I}_{i\neq j}= \frac{\chi^2_{i+,j+}+\chi^2_{i-,j-}-\chi^2_{i+,j-} -\chi^2_{i-,j+}}{4\Delta\theta_i\Delta\theta_j}
\end{equation}

where $\chi^2_{i\pm,j\pm}=\chi^2(\theta_i\pm\Delta\theta_i,\theta_j\pm\Delta\theta_j)$ and $\Delta\theta_i$ is a very small step away from that parameter's best-fit value. In our calculations we use a frequency step size of $\Delta f = 10^{-10}\mathrm{Hz}$, a normalized amplitude step size of $\Delta A = 10^{-6}$, and a phase step size of $\Delta p = 10^{-4}$.

We calculate the covariance matrix $\hat{\Sigma}$ of our best-fit model by taking the inverse of the Fisher matrix $\hat{\Sigma}= \mathcal{I}^{-1}$. We test whether $\hat{\Sigma}$ is positive definite by checking that all of its eigenvalues are positive. We take the diagonal elements of the covariance matrix to be the $1\sigma$ variance of our model parameters.

\section{Results}\label{sec:results}
We demonstrate the LASR technique by applying it to a synthetic time series and by applying it to $\delta$-Scuti KOI-976's short-cadence \emph{Kepler} photometry. We establish LASR's ability to measure the frequency, amplitude, and phase (\fap) of oscillations in a time series containing Gaussian noise, time gaps, overtone frequencies, and close frequency pairs in $\S$\ref{sec:synthetic}. We demonstrate that LASR provides a much better fit to KOI-976's complex variable signal than traditional time-domain prewhitening in $\S$\ref{sec:koi976}.

\subsection{LASR Subtraction of Synthetic Oscillations} \label{sec:synthetic}
We create a synthetic time series containing seven oscillation modes commonly seen in a $\delta$-Scuti variable star. We list the global parameters of the synthetic data in Table \ref{table:synthparams}. We create a 90-day time series with a 1-minute cadence and add Gaussian noise and data gaps that would commonly occur in \emph{Kepler} photometry. We include five data gaps: four periodic gaps that represent masked-out transits, and one large gap representing the gaps commonly seen in short-cadence \emph{Kepler} photometry. The synthetic time series includes correlated noise commonly seen in \emph{Kepler} data. We generate Gaussian noise using a Box-Muller transform \citep{box1958note,press1992random} and weight each uncertainty with a lag-1 autocorrelation between successive time samples \citep{haykin2006nonlinear}.

We include a small periodic transit in our synthetic time series to test its effects on our output code. The transit represents an Earth-radius planet orbiting a $\delta$-Scuti star with a transit depth smaller than \emph{Kepler's} detection limit. We list the transit parameters in Table \ref{table:synthparams}. When testing its effects on our fitting process, we find this injected transit causes no significant influence on our results. In general, we find that transits do not influence the LASR algorithm until their transit depths grow larger than its photometric 1-$\sigma$ uncertainty. At that limit, transits are readily visible in the time series and should be removed. {\bf We include this transit to show that low-amplitude transits at or below the detection limit do not noticeably affect out fitting results. In our algorithm, we treat removing time bins affected by transits as a standalone prerequisite before applying our significance reduction routine.}

We add seven oscillation modes to the synthetic data to test LASR's capabilities. We include a single high-amplitude oscillation at $228.7\mhz$ as an example of a stand-alone mode in the dataset. We add a close frequency pair at $252.5\mhz$ and $252.6\mhz$ to test LASR's ability to reduce oscillations that cannot be individually subtracted through spectral analysis. Additionally, we include three frequencies at integer multiples of one another to test LASR's ability to remove resonant, interdependent frequencies. We also add one oscillation whose photometric amplitude matches the synthetic data's $1\sigma$ uncertainty value to test our algorithm's ability to remove frequencies near the limit of statistical significance.

We subtract oscillations in order of highest-significance peak to lowest-significance (see Table \ref{table:synthresults} and Figure \ref{fig:synthperiodo}). LASR subtracts the single large peak ($f_1$) quickly and without difficulty. For the close frequency pair $f_2$ and $f_3$, we imitate a real time series by falsely identifying it as a single peak (see Figure \ref{fig:synthclosefreqs}). We tried a single starting frequency value of $252.53\mhz$ and could not reduce the window's significance below 67.4\%. We then set LASR to remove two close frequencies and immediately found the values listed in Table \ref{table:synthresults}. LASR also yielded accurate (\fap) values for the resonant $f_4$, $f_5$, and $f_6$ frequencies in a simultaneous fit.  

\subsection{LASR Comparison to Prewhitening: KIC 9700322}
\begin{figure}[th]
\includegraphics{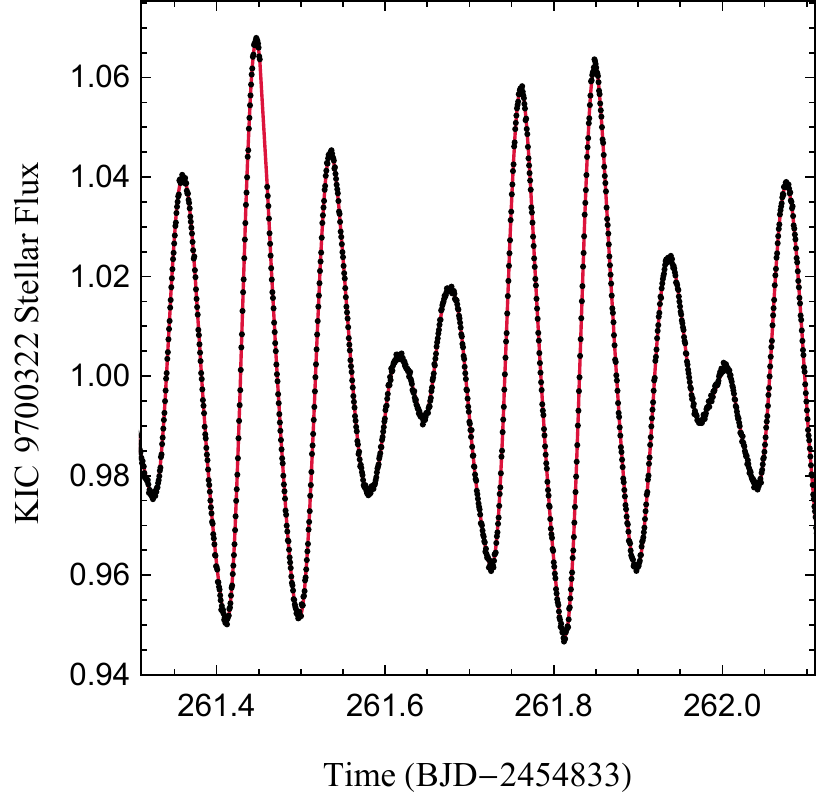}
\caption{\footnotesize Sample of KIC 9700322's stellar variability (black) and our best-fit of the variable signal (red) using LASR. Our fit yeilds a reduced $\chi^2$ value of 1.13.}
\label{fig:KIC9700322}
\end{figure}

We subtract variability from the $\delta$-Scuti star KIC 9700322 and compare our results to \citet{breger2011regularities}, who fit stellar variability using the statistical package {\tt period04}. Following \citet{breger2011regularities}, we use incorporate \emph{Kepler's} third quarter short cadence photometry in our fit and measure 76 frequencies. We show our best-fit of the stellar variability in Figure \ref{fig:KIC9700322} and compare the five highest-amplitude and five lowest-amplitude oscillations to \citet{breger2011regularities} in Table \ref{table:KIC9700322}.

\renewcommand{\arraystretch}{1.3}
\begin{table*}[htp] 
\centering
\noindent\begin{tabular}{l|llll}
\hline
\hline
$f_\mathrm{\#}$ &
\begin{tabular}{l}\hspace{-0.75em}$f_\mathrm{LASR}$ \\\hspace{-0.75em}($\mhz$) \end{tabular} & 
\begin{tabular}{l}\hspace{-0.75em}$f_\mathrm{PERIOD04}$ \\\hspace{-0.75em}($\mhz$) \end{tabular} &
\begin{tabular}{l}\hspace{-0.75em}$a_\mathrm{LASR}$ \\\hspace{-0.75em}($\mathrm{10^{-3}}$) \end{tabular} & 
\begin{tabular}{l}\hspace{-0.75em}$a_\mathrm{PERIOD04}$ \\\hspace{-0.75em}($\mathrm{10^{-3}}$) \end{tabular}  \\ \hline
1 & 145.473$\pm$1.8e-5 & 145.472 & 29.391$\pm$0.003 & 29.463 \\
2 & 113.339$\pm$2e-5 & 113.339 & 27.268$\pm$0.003 & 27.266 \\
3 & 258.811$\pm$8e-5 & 258.812 & 4.899$\pm$0.003 & 4.902 \\
4 & 290.945$\pm$0.00015 & 290.945 & 2.665$\pm$0.003 & 2.663 \\
5 & 32.1338$\pm$0.0002 & 32.133 & 2.637$\pm$0.003 & 2.633 \\
... & ... & ... & ... & ... \\
72 & 727.361$\pm$0.014 & 727.362 & 0.015$\pm$0.003  & 0.019 \\
73 & 263.59$\pm$0.02 & 263.588 & 0.014$\pm$0.003  & 0.015 \\
74 & 392.96$\pm$0.02 & 393.002 & 0.0141$\pm$0.003  & 0.015 \\
75 & 289.09$\pm$0.02 & 289.096 & 0.0135$\pm$0.003 & 0.016 \\
76 & 598.96$\pm$0.02 & 598.982 & 0.0126$\pm$0.003 & 0.014 \\
\end{tabular}
\caption{\footnotesize Stellar frequencies and amplitudes of $\delta$-Scuti KIC 9700322 measured using our algorithm and using the prewhitening program {\tt period04} \citep{breger2011regularities}. We find that LASR and prewhitening produce almost identical results with little or no discrepancy between frequency and amplitude values. \citet{breger2011regularities} lists their frequency uncertainty as $0.001\mhz$ and amplitude uncertainty as $0.003$ for all modes of oscillation.}
\label{table:KIC9700322}
\end{table*}

Our best-fit using LASR agrees almost completely with the frequencies and amplitudes found in \citet{breger2011regularities}. We successfully model KIC 9700322's stellar variability throughout \emph{Kepler's} Q3 short cadence time series. We obtain a reduced $\chi^2$ value of 1.13 for our fit; KIC 9700322's slightly reddened noise and its photometric outliers were the main causes for this value's deviation from unity. In the case of relatively straightforward variability, LASR and prewhitening are equivalently reliable, with the added bonus for LASR of only ever fitting a small parameter space at any time.

\subsection{LASR Subtraction of KOI-976}\label{sec:koi976}

\begin{figure*}[th]
\includegraphics[width=\textwidth]{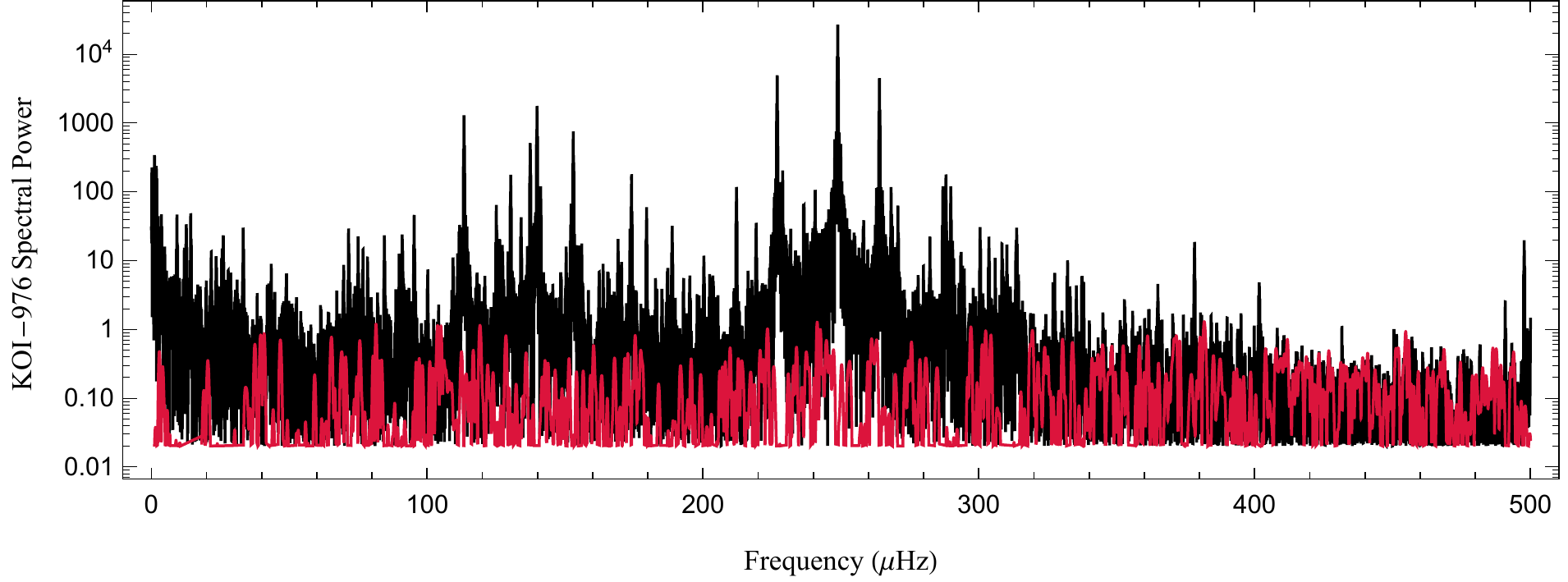}
\caption{\footnotesize Lomb-Scargle Periodogram of KOI-976's short-cadence photometry before (black) and after (red) subtracting stellar variability. LASR reduced the spectral power of all oscillations from $\sim10^4$ at the max to $\sim1.0$. We reduce the representative significance of all subtracted oscillations by at least 98\% and find that all significant oscillations are well-modelled as sinusoids.}
\label{fig:periodo}
\end{figure*}

We perform the same variability subtraction process as that described in $\S$\ref{sec:synthetic} on Kepler Object of Interest (KOI) 976, a rapidly rotating $\delta$-Scuti star that hosts an eclipsing binary companion. KOI-976 displays typical seismic activity for a $\delta$-Scuti variable, possessing a few dominant nonradial modes between $\sim100\mhz-300\mhz$, as well as many low-amplitude oscillations spanning $\sim0\mhz-500\mhz$. In this analysis, we treat the star as a rigid rotator whose variable signal is well-modelled as a linear combination of sinusoids, which typically serves as an adequate assumption for $\delta$-Scuti stars.

We perform this analysis on KOI-976's two available quarters of 1-minute \emph{Kepler} photometry available on the Mikulski Archive for Space Telescopes. We use KOI-976's presearch data conditioning data (PDC) available through the \emph{Kepler} analysis pipeline \citep{smith2012kepler} and find no relevant differences between the PDC and raw-data versions of the photometry. KOI-976's time series contains a single transit by its stellar companion, as well as several significant data gaps. We mask out the transit and treat it as another gap in the time series. These gaps produce significant aliasing in periodograms. As we show in Figure \ref{fig:synthpeak}, LASR subtraction of an oscillation removes both a peak and its aliases, so even very large data gaps are surmountable through this technique.

We follow the process detailed in $\S$\ref{sec:methods} and remove KOI-976's oscillations from the time series one at a time, except in the cases of close frequency pairs and overtone frequencies. We start with the highest-significance peak and work our way down to the significance detection limit based on KOI-976's photometric uncertainty of $\sim10^{-4}$. In total, we subtract off 319 frequencies. Figure \ref{fig:periodo} shows the original and reduced frequency power spectrum of this dataset. 

\begin{figure}[thbp]
\begin{tabular}{l}
\vspace{-1mm} \includegraphics[trim=0 35 0 0, clip]{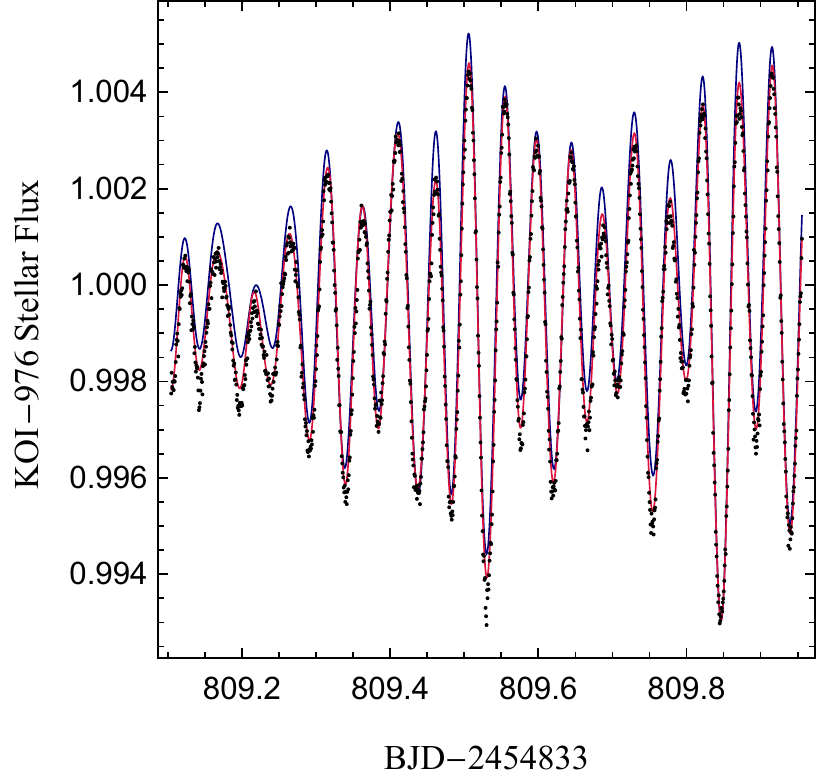} \\
\includegraphics[trim=-15.8 3 0 0, clip,width=0.4507\textwidth]{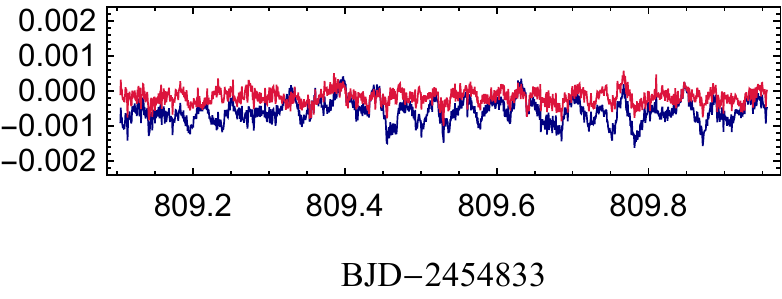}
\end{tabular}
\caption{\footnotesize A comparison of variability subtraction using LASR (red) and traditional prewhitening (blue). We fit all KOI-976 short-cadence photometry and display a $\sim1$ day sample of KOI-976's short-cadence photometry with best-fit (top) and its best-fit residuals (bottom). LASR produces a superior fit of KOI-976's seismic activity with a log-likelihood ratio of $-2\log(\mathcal{L}_{PW}/\mathcal{L}_{LASR})=33806$, which illustrates our motivation to create this technique, as it better-reduces complex seismic signals of classical pulsators.}
\label{fig:fitplot}
\end{figure}

LASR successfully minimizes all significant frequencies present in KOI-976's short-cadence photometry. Figure \ref{fig:fitplot} contrasts LASR with prewhitening and shows that LASR provides a better fit of KOI-976's oscillations than prewhitening through linear regression. We obtain a log-likelihood ratio between the two methods of $-2\log(\mathcal{L}_{PW}/\mathcal{L}_{LASR})=33806$, indicating a superior resolution of KOI-976's variability using our technique. We perform error analysis following $\S$\ref{sec:error}.




\section{Discussion \& Conclusion}\label{sec:discussion}
Our results show that LASR successfully removes stellar variability commonly seen in classical pulsators. Our technique can remove oscillations from photometry of arbitrary complexity so long as they are well-modelled as sinusoids. We find that for the rapidly-rotating $\delta$-Scuti KOI-976, LASR serves as a superior method for variability subtraction over the traditional prewhitening approach of linear regression in the time domain. In particular, we find that LASR more accurately fits the frequencies of individual oscillations. It also better-resolves close frequency pairs that can be very difficult to identify when fitting in the time domain. Combined with LASR's reliability, relatively low computation cost, and ease-of-use, we consider our technique to be a useful tool for spectral analysis in asteroseismology.

We develop LASR out of necessity: we originally attempted to subtract variability from KOI-976 following traditional prewhitening methodology that has successfully resolved the oscillations of other $\delta$-Scuti stars \citep{breger2011regularities,breger2012relationship}. We found, however, that for KOI-976, we could not obtain accurate frequencies using this technique. We observed several undesired aliasing effects occurring in the low-frequency range of our dataset due to these imperfect fits. We demonstrate that for this dataset, LASR successfully minimizes significant frequencies without producing aliasing effects (Figure \ref{fig:periodo}) and provides a better fit of KOI-976's variable signal than prewhitening (Figure \ref{fig:fitplot}).

We put forth LASR as one of many tools available for analyzing photometry. Existing tools such as Period04 \citep{lenz2004period04} and others \citep{akritas1996linear,vio2002joint,rohlfs2013tools} provide robust and well-established techniques for signal processing of photometry. Additional techniques exist for analyzing stellar variability that contains non-sinusoidal or pseudo-periodic signals. These methods include modelling signals as multivariate random variables through Gaussian processes \citep{mackay1998introduction,rasmussen2006gaussian,aigrain2016k2sc}, detecting signals through autocorrelation functions \citep{edelson1988discrete,mcquillan2014rotation}, and analyzing time-variable signals through wavelet analysis. These techniques are important tools for transit detection, denoising signals, and removing pseudo-periodic stellar signals or oscillations that vary with time. They are particularly useful for analysis of dwarf stars, where solar flares, sunspots, and non-rigid stellar rotation produce complex variable signals in photometry that must be modelled as random events. These techniques produce strong results in subtracting stellar variability, but often come with the complications of being computationally expensive or from treating stellar variability as a random. The LASR routine serves as an inexpensive and straightforward tool for analyzing high-mass stars, which typically do not possess sunspots or flares \citep{didelon1984stellar}, whose surfaces behave as rigid rotators \citep{suarez2005modelling}, and whose oscillations commonly remain constant over long timescales when on the main sequence \citep{breger1998period}.

LASR currently subtracts stellar variability that is well-modelled as a linear combination of sinusoids. Future works can expand this technique to combine LASR's significance reduction with Gaussian processes or wavelet analysis to analyze other forms of stellar variability. Such an approach could resolve a stellar signal without sacrificing information by treating seismic activity as a random process. Additionally, wavelet analysis could expand LASR's purview to the variable signal of heartbeat stars \citep{hambleton2013physics,smullen2015heartbeat} in the future.



\bibliographystyle{aa}
\bibliography{LASR_paper_arxiv}


\onecolumn
\appendix
\section{LASR Algorithm}
\subsection{Inputs and Usage}\label{app:inputs}
LASR uses inputs to control its downhill simplex routine that customize its behavior to the target time series. We list the global parameters that typically remain constant throughout the subtraction process:
\begin{enumerate}
\item Frequency scale factor to set initial downhill step sizes. When working with KOI-976's high-precision photometry, we use a scale factor of $10^{-3}\mhz$.
\item Amplitude scale factor. Within an order of magnitude of the highest-significance oscillation's amplitude  is typically adequate. This value can be set smaller as LASR subtracts smaller oscillations in the dataset.
\item Phase scale, typically set to $1.0$.
\item Number of downhill steps for LASR to take. Convergence typically occurs within 50-100 steps for an independent oscillation, but requires about twice as many steps for close frequency pairs or overtone frequencies.
\item The half-width of the peak of the highest-significance frequency in the dataset. This value determines the width of the periodogram window for LASR to sample during each subtraction. For KOI-976 we used a halfwidth of $0.35\mhz$.
\item The number of frequencies to sample in the periodogram window. We find that $10$ points provide an adequate sampling of the periodogram window in our analysis, but because of LASR's relatively low computational cost we use $25$ points.
\end{enumerate}

LASR requires the following inputs to subtract an oscillation:
\begin{enumerate}
\item Starting guess for frequency. Easily obtained via periodogram with sufficient accuracy.
\item Starting guess for amplitude. Order-of-magnitude values serve an adequate guess, as shown in Table \ref{table:synthresults}.
\item Starting guess for phase. Any value between 0 and $2\pi$ typically suffices.
\end{enumerate}

\subsection{LASR Pseudocode}\label{app:code}
We write this algorithm in {\tt c++} using established techniques for periodograms and downhill simplex routines following \citep{press2007numerical}. Our program includes proprietary optimization code and personalized libraries that make direct sharing of this program impractical. However, the LASR routine is straightforward to create. We provide an outline below that uses a periodogram window as a black-box function in a downhill simplex routine to minimize significance and determine an oscillation's (\fap) values. We also make an open-source version of our code available for download at \url{https://github.com/jpahlers/LASR}. 

\begin{algorithm}
\DontPrintSemicolon
	\KwData{$G=(T,U,\delta U)$ ~~~~time series}
	\KwResult{$G'=(T,V,\delta V)$ ~~~~time series with highest-amplitude oscillation subtracted}
	{\bf Function:} $S(G|\omega,A,\delta)$ ~~~~calculates significance in periodogram window ($\S$\ref{sec:algorithm}) centered on frequency to subtract. \;
		
	\Begin{
		{\bf Remove all time bins affected by transits or other discrete events}\;
		Set downhill simplex scale factors: \{$f_{\mathrm{sc}}$, $a_{\mathrm{sc}}$, $p_{\mathrm{sc}}$\}\;
		Set number of downhill steps to take: $N$\;
		Set number of samples in periodogram window: $npts$\;
				
		\For{$\mathrm{\#~frequencies}$}{
			Set starting guesses for of oscillation values to be fit: \{$f_{\mathrm{start}}$, $a_{\mathrm{start}}$, $p_{\mathrm{start}}$\}\;
			Calculate initial significance of periodogram window before subtraction\;
			\For{$n\in N$}{
				Take step in downhill simplex using $S(G|\omega,A,\delta)$ as black-box function \citep{press2007numerical}\;
				Update (\fap) best estimates\;			 
			}
			\For{$t\in T$}{
				Subtract oscillation from flux value: $V = U - A\sin(\omega x + p)$\;
			}
			$G \longleftarrow G'$\;
			Store best-fit (\fap) results\;
		}
		Calculate best-fit confidence intervals ($\S$\ref{sec:error})\;
		Propagate uncertainty of oscillation subtraction into reduced timeseries.\;
	}
\end{algorithm}

\newpage
\subsection{Benchmark Results}
We list benchmark performance results of the LASR routine below using the KOI-976 short-cadence data set. The computation time scales with the size of the time series and the desired precision, driven mainly by the cost of calculating periodogram significance values. To speed up our computation time, we compute trigonometric recurrences of the periodogram \citep{press2007numerical} using two processors in parallel. In general, we find that $N=100$ downhill steps is typically enough to determine the best-fit (\fap) values of an oscillation to six significant figures and that 25 periodogram window samples robustly represents a frequency's significance.   

\renewcommand{\arraystretch}{1.2}
\
\begin{table}[h]
\centering 
\begin{tabular}{l l}
\hline
\hline
{\bf Benchmark Quantity} & {\bf Value}  \\ \hline
Number of downhill steps ($N$) & 100 \\
Time series data points & 91235 \\
Periodogram window samples & 25 \\
Computation time & $12.58~\mathrm{s}$
\end{tabular}
\label{table:benchmark}
\caption{\footnotesize Quantities describing the LASR's computation time using KOI-976's short cadence photometry. Our algorithm typically fits the frequency, amplitude, and phase of a single oscillation in a timeseries in approximately 100 downhill steps that each take $\sim0.1$ when applied to KOI-976's 91235 short-cadence time bins.}
\end{table}

The computation time is primarily dedicated to calculating periodograms. \citet{press2007numerical} shows how, depending on the choice of algorithm, Lomb-Scargle periodograms scale as $N\log(N)$, where $N$ is the number of points in the time series. The fitting precision is controlled by the LASR's downhill simplex algorithm. The rate of convergence can vary largely between datasets, but we find that when fitting one oscillation (three parameters) in  KOI-976's short cadence photometry, we typically achieve four significance figures after $\sim50$ downhill steps and six significant figures after $\sim100$ downhill steps.

\section{Derivation: One Minimum Per Parameter} \label{app:deriv}
The oscillation significance $S(\omega,A,\delta)$ is a black box function that represents the significance of the residuals of a subtracted oscillation. In this section we provide a derivation showing that $S(\omega,A,\delta)$ has only one minimum per parameter (i.e. that $S(\omega,A,\delta)$ is U-shaped in each dimension over all values) so long as the actual frequency and subtracted frequency are relatively close ($|(\omega_1-\omega_2)/(\omega_1+\omega_2)|\lesssim0.1$).

We start with two sine waves $\psi_1=A_1\sin(\omega_1t+\delta_1)$ and $\psi_2=A_2\sin(\omega_2t+\delta_2)$, where $A_i$, $\omega_i$, and $\delta_i$ represent the amplitude, frequency, and phase of each function. A standard oscillation subtraction is therefore represented by,
\begin{equation}
\psi=\psi_1-\psi_2 = A_1\sin(\omega_1t+\delta_1)-A_2\sin(\omega_2t+\delta_2)
\label{eq:subtraction}
\end{equation}

Assuming $\omega_1\approx \omega_2$, equation \ref{eq:subtraction} can be expanded as,
\begin{equation}
\psi = A_1\sin(\omega_1t+\delta_1)-A_2\sin(\omega_1t+\delta_2)+A_2(\omega_1-\omega_2)t\cos(\omega_2t+\delta_2)+O((\omega_1-\omega_2)^3)
\end{equation}

Utilizing the Harmonic Addition theorem \citep[e.g.,][]{nahin2001science}, the zeroth-order terms $(A_1\sin(\omega_1t+\delta_1)-A_2\sin(\omega_1t+\delta_2))$ can be expressed as a single sine wave $A\sin(\omega_2t+\delta)$, where 
\begin{equation}
A=\sqrt{A_1^2+A_2^2-2A_1A_2\cos(\delta_1-\delta_2)}
\end{equation}

and 

\begin{equation}
\delta = \mathrm{atan}\left(\frac{A_1\cos(\delta_1)-A_2\sin(\delta_2)}{A_1\sin(\delta_1)-A_2\sin(\delta_2)} \right)
\end{equation}

Therefore, this subtraction can be represented as,
\begin{equation}
\psi = A\sin(\omega_2t+\delta)+A_2(\omega_1-\omega_2)t\cos(\omega_2t+\delta_2)+O((\omega_1-\omega_2)^3)
\label{eq:subtractfinal}
\end{equation}

Because $S(\omega,A,\delta)$ is roughly proportional to the square of the amplitude of $\psi$, minimizing Equation \ref{eq:subtractfinal} minimizes $S(\omega,A,\delta)$. The oscillation residual $\psi$ is minimized via minimizing $A$ and $\omega_1-\omega_2$. $A$ is minimized via minimizing $A_1-A_2$ and $\delta_1-\delta_2$. These are the \emph{only} minima that appear in $\psi$; therefore, only one global minimum per parameter exists.
\end{document}